\documentclass[journal,twocolumn]{IEEEtran}
\usepackage{}
\usepackage{graphicx}
\usepackage{epstopdf}
\usepackage{amssymb}
\usepackage{amsfonts}
\usepackage{amsmath}
\usepackage{algorithm}
\usepackage{algorithmic}
\usepackage{subeqnarray}
\usepackage{cases}
\usepackage{bm}
\usepackage{color}
\usepackage{subfigure,amsmath,amssymb,cite}
\usepackage{float}
\ifCLASSINFOpdf
\else
\fi
\begin{document}

\title{Power Allocation and Beamforming Design for IRS-aided Secure Directional Modulation Networks}
\author{ Rongen Dong, Feng Shu, Fuhui Zhou, Yongpeng Wu, Jiangzhou Wang, \emph{Fellow, IEEE}

\thanks{Rongen Dong and Feng Shu are with the School of Information and Communication Engineering, Hainan University, Haikou, 570228, China.}
\thanks{Fuhui Zhou is with the School of Electronic and Information Engineering, Nanjing University of Aeronautics and Astronautics, Nanjing, 210094, China.}
\thanks{Yongpeng Wu is with the Shanghai Key Laboratory of Navigation and Location Based Services, Shanghai Jiao Tong University, Minhang, Shanghai, 200240, China (Email: yongpeng.wu2016@gmail.com).}
\thanks{Jiangzhou Wang is with the School of Engineering, University of Kent, Canterbury CT2 7NT, U.K. (Email: {j.z.wang}@kent.ac.uk).}

%
}

\maketitle

\begin{abstract}
With the aim of boosting the security of the conventional directional modulation (DM) network, a secure DM network assisted by intelligent reflecting surface (IRS) is investigated in this paper. To maximize the secrecy rate (SR), we jointly optimize the power allocation (PA) factor, confidential message (CM) beamforming, artificial noise (AN) beamforming, and IRS reflected beamforming. To tackle the formulated problem, a maximizing SR with high-performance (Max-SR-HP) scheme is proposed, where the PA factor, CM beamforming, AN beamforming, and IRS phase shift matrix are derived by the derivative operation, generalized Rayleigh-Ritz, generalized power iteration, and semidefinite relaxation criteria, respectively. Given that the high complexity of the above scheme, a maximizing SR with low-complexity (Max-SR-LC) scheme is proposed, which employs the generalized leakage and successive convex approximation algorithms to derive the variables. Simulation results show that both the proposed schemes can significantly boost the SR performance, and are better than the equal PA, no IRS and random phase shift IRS schemes.
\end{abstract}
\vspace{-0.1cm}
\section{Introduction}

With the advent of the information era, the issue of communication security has gained importance. As a key issue in the security communication, the physical layer security technology has been widely investigated, such as directional modulation (DM), which has a directive property, to dramatically enhance the security of the wireless communications\cite{Daly2010Demonstration,Qiu2023Decomposed}. The key idea of DM is using techniques such as beamforming to send the confidential message (CM) to the desired direction, while simultaneously employing artificial noise (AN) to distort the signal constellation diagram\cite{Cheng2021Physical}. At present, the research on DM is centered on the radio frequency frontend and baseband. For example, by recognizing the differences and commonalities between multiple-input multiple-output and DM techniques, the authors in \cite{Ding2016MIMO} proposed a new method for the synthesis of DM transmitters, and verified the effectiveness of the proposed method with bit error rate performance metrics. In \cite{Teng2022Low}, a DM network with a malicious attacker was investigated, and three receive beamforming strategies were proposed based on maximizing the secrecy rate (SR).

Nevertheless, the rapidly growing demand for communication requires access to a huge number of devices, which leads to serious power consumption problems. Recently, the emerging intelligent reflecting surface (IRS) with low power consumption has provided a novel paradigm to tackle the issue\cite{Pan2021Reconfigurable}. IRS, made up of a large number of low-cost passive reflective elements, enables the phase shift of the incident signal to be intelligently tuned, thereby enhancing the performance of the system\cite{Yang2021Reconfigurable}. In particular, introducing IRS into DM system has been shown to significantly benefit SR performance. In \cite{Hu2020Directional}, an IRS-aided DM system was proposed, and the closed-form expression for the SR was derived. Based on the system model of \cite{Hu2020Directional}, the authors in \cite{Lin2023Enhanced} proposed two alternating iteration algorithms to further optimize the SR. Moreover, with the help of IRS, the DM network enabled the transmission of two confidential message bit streams from Alice to Bob simultaneously \cite{ShuEnhanced2021, Dong2022Low}. They all showed that optimizing the beamforming and IRS phase shift matrix can enhance the SR performance of the system.

However, the above works performed on the fixed power allocation (PA) factor. To investigate the impact of PA factor on system performance and to further boost the SR of conventional DM network, in this paper, we propose two iteration schemes that jointly optimize the PA factor, CM beamforming,  AN beamforming, and IRS phase shift matrix to maximize the SR. The main contributions of this paper are outlined as follows.

\begin{enumerate}
\item To further boost the SR performance of the conventional DM system, an IRS-assisted secure DM network is investigated. With the intention of maximizing the SR, the PA factor, beamforming vectors, and the IRS phase-shift matrix are jointly optimized. First, a maximizing SR with high-performance (Max-SR-HP) scheme is proposed. In this scheme, the closed-form expression of the PA factor is derived based on the derivative operation. The CM beamforming is derived by the generalized Rayleigh-Ritz theorem. And the AN beamforming is optimized by the generalized power iteration (GPI) algorithm. Moreover, we employe the semidefinite relaxation (SDR) strategy to design the IRS phase shift matrix.
\item Given that the high complexity of the above scheme, a maximizing SR with low-complexity (Max-SR-LC) scheme is proposed. Therein, based on the generalized leakage criterion, the closed-form expressions of the CM and AN beamforming vectors are derived by maximizing signal-to-leakage and noise ratio (Max-SLNR) and maximizing AN-to-leakage and noise ratio (Max-ANLNR) methods, respectively. And the successive convex approximation algorithm is employed to design the IRS phase shift matrix. Simulation results show that both the proposed schemes significantly outperform the benchmark schemes in terms of the SR performance.

\end{enumerate}

The remainder of this paper is organized as follows. The system model of IRS-aided secure DM network and the overall optimization problem are shown in Section \ref{s1}.
Section \ref{s2} introduces the proposed two schemes for maximizing SR.
The numerical simulation results are described in Section \ref{s3}. Finally,
the conclusions are provided in Section \ref{s4}.

{\emph{ Notations}:} in this work, we denote the scalars, vectors and matrices by lowercase, boldface lowercase, and uppercase letters, respectively. Symbols $(\cdot)^T$, $(\cdot)^H$, $\partial(\cdot)$, Tr$(\cdot)$, $(\cdot)^\dag$, $\Re\{\cdot\}$, $\text{diag}\{\cdot\}$, $|\cdot|$, and $\lambda_\text{max}\{\cdot\}$ represent the transpose, conjugate transpose, partial derivative, trace, pseudo-inverse, real part, diagonal, absolute value, and maximum eigenvalue operations, respectively. The notations $\textbf{I}_P$ and $\mathbb{C}^{P\times Q}$ refer to the identity matrix of $P\times P$ and complex-valued matrix space of $P\times Q$.

\vspace{-0.1cm}
\section{system model}\label{s1}

\begin{figure}[htbp]
\centering
\includegraphics[width=0.4\textwidth]{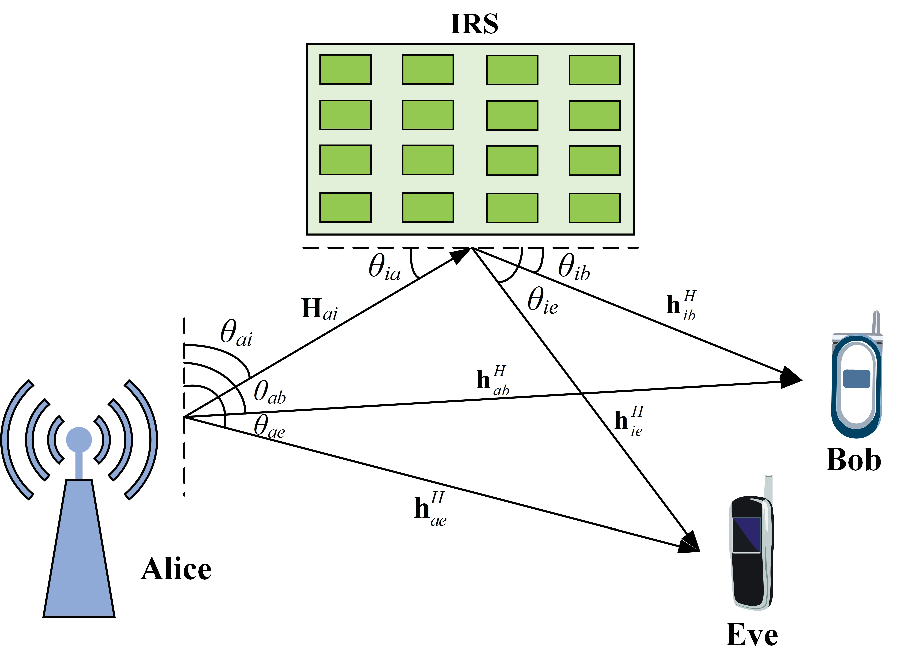}\\
\vspace{-0.3cm}
\caption{System diagram of IRS-assisted secure DM network.}\label{model}
\vspace{-0.3cm}
\end{figure}

In this paper, an IRS-aided secure DM network is considered, which is shown in Fig.~\ref{model}.  The base station (Alice) has  $N$ antennas. Both of the desired user (Bob) and eavesdropper (Eve) have single antenna. The IRS has $M$ reflective elements with adjustable phase. With the aid of the IRS, Alice sends CM to Bob while sending AN to Eve. Due to severe path loss, it is assumed that the signal reflected twice or more by the IRS is negligible. Moreover, all channels are assumed to be line-of-sight channels.
\vspace{-0.3cm}

The transmit signal at Alice is
\begin{align}
\textbf{s}_a=\sqrt{\alpha P}\textbf{v}x+\sqrt{(1-\alpha) P}\textbf{w}z,
\end{align}
where $P$ represents the total transmit power, $\alpha\in [0,1]$ and $(1-\alpha)$ stand for the PA parameters of the CM and AN, respectively. $\textbf{v}\in \mathbb{C}^{N\times1}$ and  $\textbf{w}\in \mathbb{C}^{N\times1}$ are the CM beamforming and AN beamforming, respectively, which meet $\textbf{v}^H\textbf{v}=1$ and $\textbf{w}^H\textbf{w}=1$. $x$ and $z$ stand for the CM and AN intent to Bob and Eve, respectively.

The receive signal at Bob is given by
\begin{align}\label{yb0}
y_b&=(\underbrace{\sqrt{g_{ab}}\textbf{h}^H_{ab}+\sqrt{g_{aib}}\textbf{h}^H_{ib}
\boldsymbol{\Theta}\textbf{H}_{ai}}_{\textbf{h}^H_b})\textbf{s}_a+n_b\nonumber\\
&=\sqrt{\alpha P}\textbf{h}^H_b\textbf{v}x+\sqrt{(1-\alpha) P}\textbf{h}^H_b\textbf{w}z+n_b,
\end{align}
where $g_{ab}$ and $g_{aib}=g_{ai}g_{ib}$ mean the path loss parameters of the Alice-to-Bob and Alice-IRS-Bob channels, respectively. $n_b \sim \mathcal {C}\mathcal {N} (0, \sigma^2_b)$ stands for the complex additive white Gaussian noise (AWGN), the IRS phase shift matrix $\boldsymbol{\Theta}=\text{diag}(e^{j\theta_1}, \cdots, e^{j\theta_m}, \cdots, e^{j\theta_M})$, where $\theta_m$ refers to the $m$-th phase shift element. $\textbf{h}^H_{ab}\in\mathbb{C}^{1\times N}$, $\textbf{h}^H_{ib}\in\mathbb{C}^{1\times M}$, and $\textbf{H}_{ai}=\textbf{h}_{ia}\textbf{h}_{ai}^H\in\mathbb{C}^{M\times N}$ mean the Alice-to-Bob, IRS-to-Bob, and Alice-to-IRS channels, respectively. The channel $\textbf{h}_{tr}=\textbf{h}(\theta_{tr})$, and
the normalized steering vector $\textbf{h}(\theta)$ is expressed as
\begin{align}
\textbf{h}(\theta)\buildrel \Delta \over=\frac{1}{\sqrt{K}}[e^{j2\pi\Phi_{\theta}(1)}, \cdots, e^{j2\pi\Phi_{\theta}(k)}, \cdots, e^{j2\pi\Phi_{\theta}(K)}]^T
\end{align}
where
$\Phi_{\theta}(k)\buildrel \Delta \over =-\left(k-({K+1})/{2}\right)\frac{d \cos\theta}{\lambda}, k=1, \cdots, K,$
$k$ refers to the antenna index, $d$ denotes the spacing of adjacent transmitting antennas, $\theta$ means the direction angle of departure or arrival, and $\lambda$ stands for the wavelength.

Similarly, the receive signal at Eve is
\begin{align}\label{ye0}
y_e&=(\underbrace{\sqrt{g_{ae}}\textbf{h}^H_{ae}+\sqrt{g_{aie}}\textbf{h}^H_{ie}
\boldsymbol{\Theta}\textbf{H}_{ai}}_{\textbf{h}^H_e})\textbf{s}_a+n_e\nonumber\\
&=\sqrt{\alpha P}\textbf{h}^H_e\textbf{v}x+\sqrt{(1-\alpha) P}\textbf{h}^H_e\textbf{w}z+n_e,
\end{align}
where $g_{ae}$ and $g_{aie}=g_{ai}g_{ie}$ refer to the path loss parameters of the Alice-to-Eve and Alice-IRS-Eve channels, respectively. $n_e \sim \mathcal {C}\mathcal {N} (0, \sigma^2_e)$ represents the AWGN, $\textbf{h}^H_{ae}\in\mathbb{C}^{1\times N}$ and $\textbf{h}^H_{ie}\in\mathbb{C}^{1\times M}$ stand for the Alice-to-Eve and IRS-to-Eve channels, respectively.

Based on (\ref{yb0}) and (\ref{ye0}), the achievable rates at Bob and Eve are
\vspace{-0.1cm}
\begin{align}\label{Rb0}
&R_b=\text{log}_2\Big(1+\frac{\alpha P |\textbf{h}^H_b\textbf{v}|^2}
{(1-\alpha)P|\textbf{h}^H_b\textbf{w}|^2+\sigma^2_b}\Big),
\end{align}
and
\begin{align}
&R_e=\text{log}_2\Big(1+\frac{\alpha P |\textbf{h}^H_e\textbf{v}|^2}
{(1-\alpha)P|\textbf{h}^H_e\textbf{w}|^2+\sigma^2_e}\Big),
\end{align}
respectively.

In this paper, we aim to maximize the SR by jointly optimizing the PA factor $\alpha$, CM beamforming $\textbf{v}$, AN beamforming $\textbf{w}$, and IRS phase shift matrix $\boldsymbol{\Theta}$. The overall optimization problem is formulated as follows
\begin{subequations}\label{p0}
\begin{align}
&\max \limits_{\alpha, \textbf{v}, \textbf{w}, \boldsymbol{\Theta}}
~~R_s=R_b-R_e\\
&~~~\text{s.t.} ~~~~~0\leq \alpha\leq 1, \textbf{v}^H\textbf{v}=1, \textbf{w}^H\textbf{w}=1,\\
& ~~~~~~~~~~~ |\boldsymbol{\Theta}(m,m)|=1, m=1,\dots,M.\label{theta00}
\end{align}
\end{subequations}
Noting that this is a non-convex optimization problem since the objective function and constraints are non-convex, and the variables are coupled with each other, which makes it difficult to be addressed directly. In what follows, the alternating optimization strategy is taken into account to solve the problem.

\vspace{-0.2cm}
\section{Proposed solutions}\label{s2}
In this section, two schemes, called Max-SR-HP and Max-SR-LC, are proposed to tackle the optimization problem (\ref{p0}). Therein, the alternating optimization strategy is applied to successively design the PA factor $\alpha$, CM beamforming $\textbf{v}$, AN beamforming $\textbf{w}$, and IRS phase shift matrix $\boldsymbol{\Theta}$.

First, we design the PA factor $\alpha$ by fixing $\textbf{v}$, $\textbf{w}$, and $\boldsymbol{\Theta}$. Defining $a=P|\textbf{h}_b^H\textbf{v}|^2-P|\textbf{h}_b^H\textbf{w}|^2$, $b=P|\textbf{h}_b^H\textbf{w}|^2+\sigma_b^2$, $c=P|\textbf{h}_b^H\textbf{w}|^2$, $d=P|\textbf{h}_e^H\textbf{w}|^2$, $e=d+\sigma_e^2$, $f=P|\textbf{h}_e^H\textbf{v}|^2-d$. Then, the problem (\ref{p0}) can be degenerated into the optimization problem on $\alpha$ as follows
\vspace{-0.1cm}
\begin{subequations}\label{alpha0}
\begin{align}
&\max \limits_{\alpha}
~~f(\alpha)=\frac{(a\alpha+b)(-d\alpha+e)}
{(-c\alpha+b)(f\alpha+e)}\\
&~~\text{s.t.} ~~~0\leq \alpha\leq 1.
\end{align}
\end{subequations}
We simplify the objective function as
\begin{align}\label{PD}
f(\alpha)=\frac{ad\alpha^2+(bd-ae)\alpha-be}{cf\alpha^2+(ce-bf)\alpha-be}.
\end{align}
Given that $b/c>1$, $-e/f>1$, and the denominator is not equal to 0, taking a partial derivative of (\ref{PD}) and setting it equal to 0 yields
\vspace{-0.1cm}
\begin{align}
\frac{\partial f(\alpha)}{\partial \alpha}=\frac{A\alpha^2+B\alpha+C}{(cf\alpha^2+(ce-bf)\alpha-be)^2}=0,
\end{align}
where $A=ad(ce-bf)-cf(bd-ae)$, $B=2(becf-adbe)$, $C=(ce-bf)be-(bd-ae)be$. Further simplification yields
\begin{align}
A\alpha^2+B\alpha+C=0.
\end{align}
If $A=0$, a candidate solution can be obtained as $\alpha=-C/B$. Otherwise, by using the quadratic root formula, the two candidate solutions can be found as $\alpha=(-B\pm\sqrt{B^2-4AC})/(2A)$. Then, one needs to determine whether these candidate solutions are in the interval $[0,1]$. And if so, compare their values $f(\alpha)$ with the endpoint values $f(0)$ and $f(1)$, and then obtain the optimal solution.

\vspace{-0.1cm}
\subsection{Proposed high-performance algorithm}
Given that the PA factor $\alpha$ has been designed, in this section, we turn our attention to deriving the beamforming vectors $\textbf{v}$ and $\textbf{w}$, and IRS phase shift matrix $\boldsymbol{\Theta}$.

First, we consider to optimize $\textbf{v}$. Given $\alpha$, $\textbf{w}$, and $\boldsymbol{\Theta}$, we extract $\textbf{v}$ before addressing the problem. Thus, the optimization problem regarding $\textbf{v}$ can be simplified as
\begin{align}\label{v1}
&\max \limits_{\textbf{v}}
~~\frac{\textbf{v}^H\textbf{A}\textbf{v}}
{\textbf{v}^H\textbf{B}\textbf{v}}  ~~~~\text{s.t.} ~\textbf{v}^H\textbf{v}=1,
\end{align}
\vspace{-0.15cm}
where
\vspace{-0.1cm}
\begin{align}
&\textbf{A}=\alpha P \textbf{h}_b\textbf{h}_b^H+[(1-\alpha)P|\textbf{h}^H_b\textbf{w}|^2+\sigma^2_b]\textbf{I}_N,\\
&\textbf{B}=\alpha P \textbf{h}_e\textbf{h}_e^H+[(1-\alpha)P|\textbf{h}_e^H\textbf{w}|^2+\sigma^2_e]\textbf{I}_N.
\vspace{-0.2cm}
\end{align}
Based on the generalized Rayleigh-Ritz theorem, the optimal beamforming $\textbf{v}$ is obtained by solving the eigenvector corresponding to the largest eigenvalue of the matrix $\textbf{B}^{-1}\textbf{A}$.

Then, with other variables being fixed, by ignoring the constant terms, the optimization problem with respect to beamforming $\textbf{w}$ can be degraded to
\begin{align}\label{w1}
&\max \limits_{\textbf{w}}
~~\frac{\textbf{w}^H\textbf{Q}_1\textbf{w}}
{\textbf{w}^H\textbf{Q}_2\textbf{w}}\cdot
\frac{\textbf{w}^H\textbf{Q}_3\textbf{w}}
{\textbf{w}^H\textbf{Q}_4\textbf{w}}   ~~~~\text{s.t.} ~\textbf{w}^H\textbf{w}=1,
\end{align}
where
\begin{subequations}
\begin{align}
&\textbf{Q}_1=(1-\alpha)P\textbf{h}_b\textbf{h}_b^H+(\alpha P |\textbf{h}_b^H\textbf{v}|^2+\sigma^2_b)\textbf{I}_N,\\
&\textbf{Q}_2=(1-\alpha)P\textbf{h}_b\textbf{h}_b^H+\sigma^2_b\textbf{I}_N,\\
&\textbf{Q}_3=(1-\alpha)P\textbf{h}_e\textbf{h}_e^H+\sigma^2_e\textbf{I}_N,\\
&\textbf{Q}_4=(1-\alpha)P\textbf{h}_e\textbf{h}_e^H+(\alpha P|\textbf{h}^H_e\textbf{v}|^2+\sigma^2_e)\textbf{I}_N.
\end{align}
\end{subequations}
At this point, in accordance with the GPI algorithm in \cite{LEE2008ICC}, the optimal $\textbf{w}$ can be derived.

Next, when fixing $\alpha$, $\textbf{v}$, and $\textbf{w}$, we define that
\vspace{-0.1cm}
\begin{align}
&\boldsymbol{\theta}=[e^{j\theta_1}, \cdots, e^{j\theta_m}, \cdots, e^{j\theta_M}]^H, \widetilde{\boldsymbol{\theta}}=\left[\boldsymbol{\theta}; 1\right],\\
&\textbf{h}_{b1}=\left[ \begin{array}{*{20}{c}}
\text{diag}\{\sqrt{\alpha Pg_{aib}}\textbf{h}^H_{ib}\}\textbf{H}_{ai}\textbf{v}\\
\sqrt{\alpha Pg_{ab}}\textbf{h}^H_{ab}\textbf{v}
\end{array}\right]_{(M+1)\times 1},  \\
&\textbf{h}_{b2}=\left[ \begin{array}{*{20}{c}}
\text{diag}\{\sqrt{(1-\alpha)Pg_{aib}}\textbf{h}^H_{ib}\}\textbf{H}_{ai}\textbf{w}\\
\sqrt{(1-\alpha) Pg_{ab}}\textbf{h}^H_{ab}\textbf{w}
\end{array}\right]_{(M+1)\times 1},\\
&\textbf{h}_{e1}=\left[ \begin{array}{*{20}{c}}
\text{diag}\{\sqrt{\alpha Pg_{aie}}\textbf{h}^H_{ie}\}\textbf{H}_{ai}\textbf{v}\\
\sqrt{\alpha Pg_{ae}}\textbf{h}^H_{ae}\textbf{v}
\end{array}\right]_{(M+1)\times 1},\\
&\textbf{h}_{e2}=\left[ \begin{array}{*{20}{c}}
\text{diag}\{\sqrt{(1-\alpha) Pg_{aie}}\textbf{h}^H_{ie}\}\textbf{H}_{ai}\textbf{w}\\
\sqrt{(1-\alpha) Pg_{ae}}\textbf{h}^H_{ae}\textbf{w}
\end{array}\right]_{(M+1)\times 1}.
\end{align}
Then, the objective function of problem (\ref{p0}) can be recast as
\begin{align}\label{function}
R_s&=\text{log}_2\Big(1+\frac{|\widetilde{\boldsymbol{\theta}}^H\textbf{h}_{b1}|^2}
{|\widetilde{\boldsymbol{\theta}}^H\textbf{h}_{b2}|^2+\sigma_b^2}\Big)-
\text{log}_2\Big(1+\frac{|\widetilde{\boldsymbol{\theta}}^H\textbf{h}_{e1}|^2}
{|\widetilde{\boldsymbol{\theta}}^H\textbf{h}_{e2}|^2+\sigma_e^2}\Big)\nonumber\\
&=\text{log}_2\Big(\frac{|\widetilde{\boldsymbol{\theta}}^H\textbf{h}_{b1}|^2+
|\widetilde{\boldsymbol{\theta}}^H\textbf{h}_{b2}|^2+\sigma_b^2}
{|\widetilde{\boldsymbol{\theta}}^H\textbf{h}_{b2}|^2+\sigma_b^2}\Big)+\nonumber\\
&~~~~\text{log}_2\Big(\frac{|\widetilde{\boldsymbol{\theta}}^H\textbf{h}_{e2}|^2+\sigma_e^2}
{|\widetilde{\boldsymbol{\theta}}^H\textbf{h}_{e1}|^2+|\widetilde{\boldsymbol{\theta}}^H\textbf{h}_{e2}|^2+
\sigma_e^2}\Big)\nonumber\\
&=\text{log}_2\Bigg(\frac{\widetilde{\boldsymbol{\theta}}^H[\overbrace{\textbf{h}_{b1}\textbf{h}_{b1}^H+
\textbf{h}_{b2}\textbf{h}_{b2}^H+1/(M+1)\sigma_b^2\textbf{I}_{M+1}}^{\textbf{B}_1}]\widetilde{\boldsymbol{\theta}}}
{\widetilde{\boldsymbol{\theta}}^H[\underbrace{\textbf{h}_{b2}\textbf{h}_{b2}^H+1/(M+1)\sigma_b^2\textbf{I}_{M+1}}_{\textbf{B}_2}]
\widetilde{\boldsymbol{\theta}}}\Bigg)+\nonumber\\
&~~~\text{log}_2\Bigg(\frac{\widetilde{\boldsymbol{\theta}}^H[\overbrace{
\textbf{h}_{e2}\textbf{h}_{e2}^H+1/(M+1)\sigma_e^2\textbf{I}_{M+1}}^{\textbf{E}_1}]\widetilde{\boldsymbol{\theta}}}
{\widetilde{\boldsymbol{\theta}}^H[\underbrace{\textbf{h}_{e1}\textbf{h}_{e1}^H+
\textbf{h}_{e2}\textbf{h}_{e2}^H+1/(M+1)\sigma_e^2\textbf{I}_{M+1}}_{\textbf{E}_2}]
\widetilde{\boldsymbol{\theta}}}\Bigg).
\vspace{-1cm}
\end{align}
Given that the objective function (\ref{function}) is non-convex, to facilitate handling, we regard the relaxation of the optimization problem (\ref{p0}) with respect to $\boldsymbol{\Theta}$ into a semidefinite programming problem. Defining $\widetilde{\boldsymbol{\Theta}}=\widetilde{\boldsymbol{\theta}}\widetilde{\boldsymbol{\theta}}^H$, the optimization problem with respect to $\boldsymbol{\Theta}$  is given by
\begin{subequations}\label{Theta0}
\begin{align}
&\max \limits_{\widetilde{\boldsymbol{\Theta}}}
~~\text{log}_2(\text{Tr}(\textbf{B}_1\widetilde{\boldsymbol{\Theta}}))+
\text{log}_2(\text{Tr}(\textbf{E}_1\widetilde{\boldsymbol{\Theta}}))-\nonumber\\
&~~~~~~~~~\text{log}_2(\text{Tr}(\textbf{B}_2\widetilde{\boldsymbol{\Theta}}))-
\text{log}_2(\text{Tr}(\textbf{E}_2\widetilde{\boldsymbol{\Theta}}))
\\
&~~\text{s.t.} ~~~\widetilde{\boldsymbol{\Theta}}\succeq 0, \text{rank}(\widetilde{\boldsymbol{\Theta}})=1,\\
&~~~~~~~~\widetilde{\boldsymbol{\Theta}}(m,m)=1, m=1, \dots M+1.
\end{align}
\end{subequations}
In accordance with the Majorization-Minimization algorithm in \cite{Sun2017MM}, at any fixed point $\widehat{\textbf{Y}}$, we have
\vspace{-0.05cm}
\begin{align}\label{MM}
\text{In}(\text{Tr}(\textbf{X}\textbf{Y}))\leq \text{In}(\text{Tr}(\textbf{X}\widehat{\textbf{Y}}))+
{\text{Tr}(\textbf{X}(\textbf{Y}-\widehat{\textbf{Y}}))}/{\text{In}(\text{Tr}(\textbf{X}\widehat{\textbf{Y}}))}.
\vspace{-0.05cm}
\end{align}
By applying (\ref{MM}) to (\ref{Theta0}) and eliminating the rank-one constraint, (\ref{Theta0}) can be recast as
\begin{subequations}\label{Theta2}
\begin{align}
&\max \limits_{\widetilde{\boldsymbol{\Theta}}}
~~\text{In}(\text{Tr}(\textbf{B}_1\widetilde{\boldsymbol{\Theta}}))+
\text{In}(\text{Tr}(\textbf{E}_1\widetilde{\boldsymbol{\Theta}}))-\nonumber\\
&~~~~~~~~~\text{Tr}((\textbf{B}_2/\text{Tr}(\textbf{B}_2\widehat{\boldsymbol{\Theta}})+
\textbf{E}_2/\text{Tr}(\textbf{E}_2\widehat{\boldsymbol{\Theta}}))\widetilde{\boldsymbol{\Theta}})
\\
&~~\text{s.t.} ~~~\widetilde{\boldsymbol{\Theta}}\succeq 0, \widetilde{\boldsymbol{\Theta}}(m,m)=1, m=1, \dots M+1.
\end{align}
\end{subequations}
It is a convex optimization problem and can be addressed directly with convex optimization tools such as CVX. Since the solution computed by the SDR algorithm fails to guarantee the rank-one property, Gaussian randomization is applied to recover the rank-one of the problem (\ref{Theta2}).

The overall processing of the proposed Max-SR-HP scheme is as follows: first, we initialize the PA factor $\alpha$, CM beamformnig $\textbf{v}$, AN beamforming $\textbf{w}$, and IRS phase shift matrix $\boldsymbol{\Theta}$ to feasible solutions. Then, update $\alpha$, $\textbf{v}$, $\textbf{w}$, and $\boldsymbol{\Theta}$ sequentially by solving for problems (\ref{alpha0}), (\ref{v1}), (\ref{w1}), and (\ref{Theta2}). The alternating iteration process is repeated among the four variables until the termination condition is satisfied, i.e., $|R_s^{(t)}-R_s^{(t-1)}|\leq \epsilon$, where $t$ and $\epsilon$ mean the iteration number and convergence accuracy, respectively. Moreover, the overall computational complexity of this scheme is $\mathcal {O}(L_{HP}(N^3+2N^2M+(N^3+4N^2M^2)\text{In}(1/\epsilon)+(\sqrt{2}(M+1)^{4.5}+JM^3)\text{In}(1/\epsilon)))$  float-point operations (FLOPs), where $L_{HP}$ and $J$ are the overall number of iterations and number of randomizations, respectively.

\vspace{-0.25cm}
\subsection{Proposed low-complexity algorithm}

Given that the high computational complexity of the above scheme, the Max-SR-LC scheme with low-complexity is developed to tackle the problem (\ref{p0}). Specifically, the Max-SLNR and Max-ANLNR strategies are employed to design the CM beamforming $\textbf{v}$ and AN beamforming $\textbf{w}$, respectively, and the SCA method is applied to optimize IRS phase shift matrix $\boldsymbol{\Theta}$.

Fixing $\alpha$, $\textbf{w}$, and $\boldsymbol{\Theta}$, the channels $\textbf{h}^H_{ab}$, $\textbf{H}_{ai}$, and $\textbf{h}^H_{ib}$ are viewed as the desired channels, while $\textbf{h}^H_{ae}$ is viewed as the undesired channel. Based on the leakage criterion in \cite{Sadek2007A}, the CM beamforming $\textbf{v}$ can be found by solution of the Max-SLNR problem as follows
\begin{align}
&\max \limits_{\textbf{v}}
~~\text{SLNR}=\frac{\text{Tr}\{\textbf{v}^H\textbf{F}_1\textbf{v}\}}
{\text{Tr}\{\textbf{v}^H\textbf{F}_2\textbf{v}\}}   ~~~~\text{s.t.} ~\textbf{v}^H\textbf{v}=1,
\end{align}
\vspace{-0.1cm}
where
\begin{align}
&\textbf{F}_1=\alpha P(g_{aib}\textbf{H}_{ai}^H\boldsymbol{\Theta}^H\textbf{h}_{ib}\textbf{h}^H_{ib}\boldsymbol{\Theta}\textbf{H}_{ai}+
g_{ab}\textbf{h}_{ab}\textbf{h}^H_{ab}),\\
&\textbf{F}_2=\alpha Pg_{ae}\textbf{h}_{ae}\textbf{h}_{ae}^H+\sigma_b^2\textbf{I}_N.
\end{align}
According to the generalized Rayleigh-Ritz theorem, the optimal beamforming $\textbf{v}$ is obtained by deriving the eigenvector corresponding to the largest eigenvalue of the matrix $\textbf{F}_2^{-1}\textbf{F}_1$.

Similarly, given $\alpha$, $\textbf{v}$, and $\boldsymbol{\Theta}$, the AN beamforming $\textbf{w}$ can be derived by resolving the Max-ANLNR problem
\begin{align}
&\max \limits_{\textbf{w}}
~~\text{ANLNR}=\frac{\text{Tr}\{\textbf{w}^H\textbf{G}_1\textbf{w}\}}
{\text{Tr}\{\textbf{w}^H\textbf{G}_2\textbf{w}\}}    ~~~~\text{s.t.} ~\textbf{w}^H\textbf{w}=1,
\end{align}
\vspace{-0.1cm}
where
\begin{align}
\textbf{G}_1=&(1-\alpha)Pg_{ae}\textbf{h}_{ae}\textbf{h}_{ae}^H,\\
\textbf{G}_2=&(1-\alpha)P(g_{aib}\textbf{H}^H_{ai}\boldsymbol{\Theta}^H\textbf{h}_{ib}
\textbf{h}^H_{ib}\boldsymbol{\Theta}\textbf{H}_{ai}+g_{ab}\textbf{h}_{ab}\textbf{h}^H_{ab})\nonumber\\
&+\sigma_e^2\textbf{I}_N.
\end{align}
Then, the optimal $\textbf{w}$ is obtained by deriving the eigenvector corresponding to the largest eigenvalue of the matrix $\textbf{G}_2^{-1}\textbf{G}_1$.

In what follows, we focus on designing the IRS phase shift matrix $\boldsymbol{\Theta}$. Fixing $\alpha$, $\textbf{v}$, and $\textbf{w}$, according to (\ref{function}), the objective function with respect to $\boldsymbol{\Theta}$ can be rewritten as $\text{log}_2(\widetilde{\boldsymbol{\theta}}^H\textbf{B}_1\widetilde{\boldsymbol{\theta}}/
(\widetilde{\boldsymbol{\theta}}^H\textbf{B}_2\widetilde{\boldsymbol{\theta}}))+
\text{log}_2(\widetilde{\boldsymbol{\theta}}^H\textbf{E}_1\widetilde{\boldsymbol{\theta}}/
(\widetilde{\boldsymbol{\theta}}^H\textbf{E}_2\widetilde{\boldsymbol{\theta}}))$. Notice that this objective function is non-convex and further transformation is required. From \cite{Moon1999Mathematical}, i.e.,
\begin{align}
\frac{\textbf{u}^H\textbf{X}\textbf{u}}{\textbf{u}^H\textbf{Y}\textbf{u}}\geq \frac{2\Re\{\overline{\textbf{u}}^H\textbf{X}\textbf{u}\}}{\overline{\textbf{u}}^H\textbf{Y}\overline{\textbf{u}}}
-\frac{\overline{\textbf{u}}^H\textbf{X}\overline{\textbf{u}}}{(\overline{\textbf{u}}^H\textbf{Y}\overline{\textbf{u}})^2}\textbf{u}^H\textbf{Y}\textbf{u},
\end{align}
and the lemma in \cite{Sun2017MM}, that is,
\begin{align}
\textbf{u}^H\textbf{Y}\textbf{u}\leq \textbf{u}^H\textbf{W}\textbf{u}+2\Re\{\textbf{u}^H(\textbf{Y}-\textbf{W})\overline{\textbf{u}}\}+\overline{\textbf{u}}^H(\textbf{W}-\textbf{Y})\overline{\textbf{u}},
\end{align}
where $\overline{\textbf{u}}$ is a feasible solution and $\textbf{W}=\lambda_{\max}(\textbf{Y})\textbf{I}$,
the objective function with respect to $\boldsymbol{\Theta}$ can be reformulated as
\begin{align}
\text{log}_2(2\Re\{\textbf{f}^H_B\widetilde{\boldsymbol{\theta}}\}+\xi_B)+
\text{log}_2(2\Re\{\textbf{f}^H_E\widetilde{\boldsymbol{\theta}}\}+\xi_E),
\end{align}
where
\vspace{-0.2cm}
\begin{align}
&\textbf{f}_K=\Big(\frac{\textbf{K}_1}{\overline{\boldsymbol{\theta}}^H\textbf{K}_2\overline{\boldsymbol{\theta}}}-
\frac{(\textbf{K}_2-\lambda_{\text{max}}(\textbf{K}_2)\textbf{I}_{M+1})\overline{\boldsymbol{\theta}}^H\textbf{K}_1\overline{\boldsymbol{\theta}}}
{(\overline{\boldsymbol{\theta}}^H\textbf{K}_2\overline{\boldsymbol{\theta}})^2}\Big)\overline{\boldsymbol{\theta}}, \\
&\xi_K=\frac{\overline{\boldsymbol{\theta}}^H\textbf{K}_1\overline{\boldsymbol{\theta}}}
{\overline{\boldsymbol{\theta}}^H\textbf{K}_2\overline{\boldsymbol{\theta}}}
-\frac{\lambda_{\text{max}}(\textbf{K}_2)2(M+1)\overline{\boldsymbol{\theta}}^H\textbf{K}_1\overline{\boldsymbol{\theta}}}
{(\overline{\boldsymbol{\theta}}^H\textbf{K}_2\overline{\boldsymbol{\theta}})^2}, K=B,E,
\vspace{-2cm}
\end{align}
and $\overline{\boldsymbol{\theta}}$ is the solution obtained in the previous iteration of $\widetilde{\boldsymbol{\theta}}$.
Moreover, the constraint (\ref{theta00}) can be relaxed to be convex $|\widetilde{\boldsymbol{\theta}}(m)|\leq 1$\cite{Shu2022Beamforming}. And the logarithmic function is monotonically increasing. Then, the optimization problem with respect to $\boldsymbol{\Theta}$ can be formulated as
\begin{subequations}\label{theta1}
\begin{align}
&\max \limits_{\widetilde{\boldsymbol{\theta}}}
~~\text{log}_2(2\Re\{\textbf{f}^H_B\widetilde{\boldsymbol{\theta}}\}+\xi_B)+
\text{log}_2(2\Re\{\textbf{f}^H_E\widetilde{\boldsymbol{\theta}}\}+\xi_E)\\
&~~\text{s.t.} ~~~~|\widetilde{\boldsymbol{\theta}}(m)|\leq 1, m=1,\dots,M,\\
&~~~~~~~~~|\widetilde{\boldsymbol{\theta}}(m)|=1, m=M+1.
\end{align}
\end{subequations}
At this point, the problem (\ref{theta1}) is convex, which can be directly addressed by the mathematical solver (e.g. CVX).

Similar to the iterative process of the proposed Max-SR-HP scheme, the proposed Max-SR-LC scheme is updated alternately among the four optimization variables until the termination condition is reached. In addition, the overall computational complexity of this scheme is $\mathcal {O}(L_{LC}(2N^3+4N^2M^2+(N^2(M+1)^2+N(M+1))\text{In}(1/\epsilon)))$ FLOPs, where $L_{LC}$ means the whole number of iterations.

 \vspace{-0.2cm}
\section{Simulation Results}\label{s3}

In this section, we verify the effectiveness of the proposed two schemes with simulation results. The default simulation parameters of the system are set as follows: $P=30$dBm, $N=8$, $M=128$, $\theta_{ai}=17\pi/36$, $\theta_{ab}=\pi/2$, $\theta_{ae}=11\pi/18$, $d_{ai}=120$m, $d_{ab}=d_{ae}=150$m, $\sigma_b^2=\sigma_e^2=-40$dBm. The path loss model is chosen to be $g_{tr}=\lambda^2/(4\pi d_{tr})^2$, where $\lambda$ and $d_{tr}$ mean the wavelength and distance. For convenience of calculation, we set $(\lambda/(4\pi))^2=10^{-2}$.

\begin{figure*}
 \setlength{\abovecaptionskip}{-5pt}
 \setlength{\belowcaptionskip}{-10pt}
 \centering
 \begin{minipage}[t]{0.33\linewidth}
  \centering
  \includegraphics[width=2.56in]{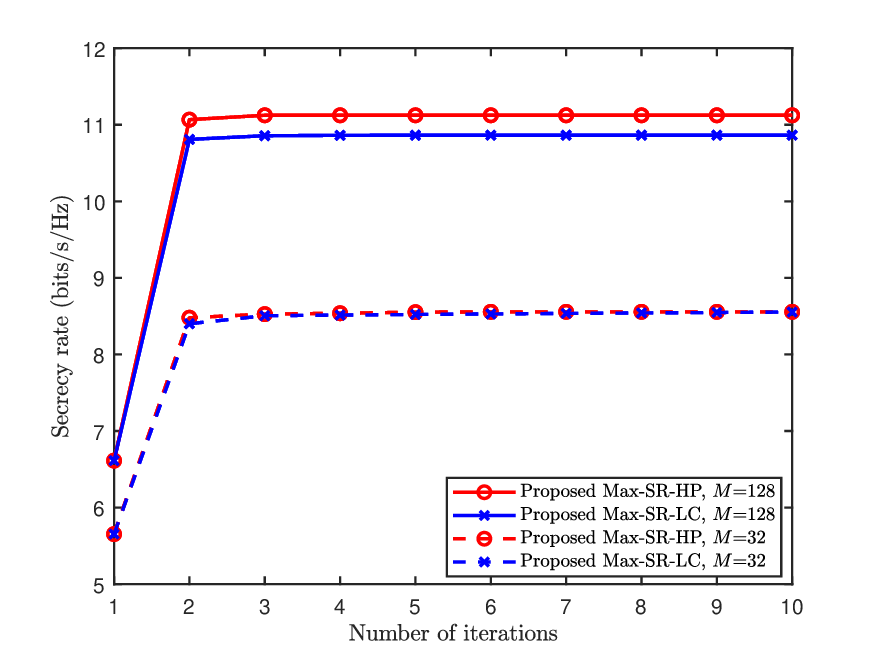}
  \caption{Convergence of proposed schemes.}\label{iter}
 \end{minipage}%
 \begin{minipage}[t]{0.33\linewidth}
  \centering
  \includegraphics[width=2.56in]{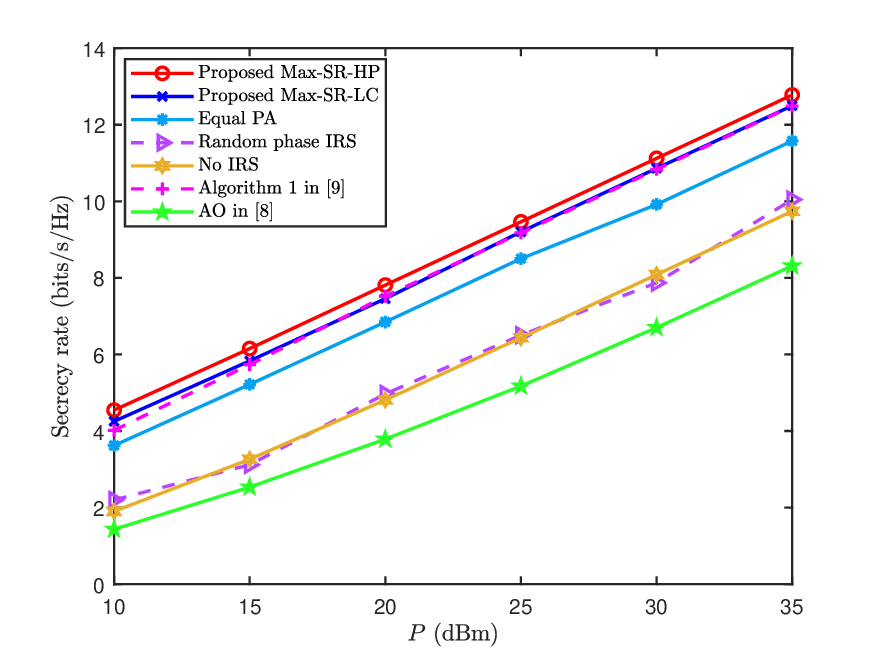}
  \caption{SR versus the transmit power $P$ of Alice.}\label{P}
 \end{minipage}
 \begin{minipage}[t]{0.33\linewidth}
  \centering
  \includegraphics[width=2.56in]{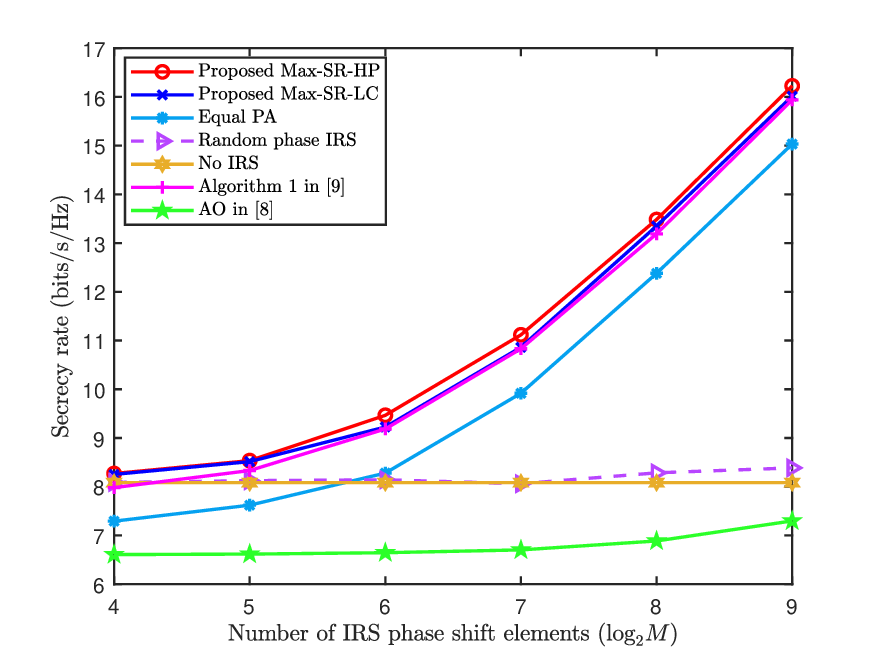}
  \caption{SR versus the number $M$ of IRS elements.}\label{M}
 \end{minipage}
 \vspace{-0.5cm}
\end{figure*}

Fig.~{\ref{iter}} depicts the curves of SR versus the number of iterations of two proposed schemes. It can be found that both the proposed schemes converge after about four iterations, regardless of the value of the number $M$ of IRS phase shift elements. The SRs of the two proposed schemes increase gradually as $M$ increases, and the SR of the proposed Max-SR-HP scheme is slightly better than that of the proposed Max-SR-LC scheme. Considering the complexity analysis in Section \ref{s2}, it is clear that the latter's low complexity is traded off with the expense of some performance loss.

Fig.~{\ref{P}} shows the SR versus the transmit power $P$ of Alice ranging from 10$\sim$35dBm. It is seen that the SRs of all schemes increase gradually with $P$. When $P=15$dBm, the SRs of the proposed Max-SR-HP scheme is 7.3\%, 17.9\%, 88.7\%, 97.2\%, and 142.7\% higher than those of the Algorithm 1 in \cite{Lin2023Enhanced}, equal PA (i.e., $\beta=0.5$), no IRS, random phase shift IRS, and alternating optimization (AO) in \cite{Hu2020Directional} schemes. In particular, as $P$ increases, the SRs of the proposed Max-SR-LC and Algorithm 1 in \cite{Lin2023Enhanced} $(\mathcal {O}(N^2(M+1)^3)\text{In}(1/\varsigma))$ are similar when $P\geq 20$dBm, while the former has a lower computational complexity since the highest order of its complexity is 2 less than the latter's 3.

Fig.~{\ref{M}} plots the SR versus the number $M$ of IRS phase shift elements. From the figure, it can be seen that the SRs of all the schemes, except for random phase shift IRS and no IRS, increase gradually with $M$. This is owing to the fact that when the IRS phase shift matrix is optimally designed, the IRS can provide more performance gain to the system as $M$ increases.
The SR of the equal PA scheme is lower than that of the no IRS scheme when $M$ is small. This is because Eve eavesdrops less CM and does not need to send more AN to interfere against Eve. Whereas AN becomes more important as $M$ gradually increases.
Moreover, regardless of the value of $M$, both the proposed schemes outperform the remaining ones in terms of the SR performance. These verify the benefits of optimized PA factor and beamforming.

\vspace{-0.5cm}
\section{Conclusion}\label{s4}

In this paper, an IRS-assisted secure DM network was investigated. To tackle the formulated problem of maximizing the SR, two schemes that jointly optimize the PA factor, CM beamforming, AN beamforming, and IRS phase shift matrix were proposed. First, the closed-form expression of the PA factor was derived with the derivative operation. Then, the proposed Max-SR-HP scheme employed the generalized Rayleigh-Ritz, GPI, and SDR criterion to derive the CM beamforming, AN beamforming, and IRS phase shift matrix, respectively. The proposed Max-SR-LC scheme utilized Max-SLNR, Max-ANLNR, and SCA methods to optimize them. Simulation results showed that both the proposed schemes are better than the existing ones with no PA optimization, and they also outperform the equal PA, no IRS, and random phase shift IRS schemes in terms of the SR performance.


\ifCLASSOPTIONcaptionsoff
  \newpage
\fi

\bibliographystyle{IEEEtran}

\bibliography{IEEEfull,reference}
\end{document}